\documentclass[a4paper,11pt,oneside]{article}

\usepackage[english]{babel}
\usepackage[T1]{fontenc}
\usepackage[utf8]{inputenc}

\usepackage[pdftex,unicode]{hyperref}
\hypersetup{pdftitle=Bertram's Pairs Trading Strategy with Bounded Risk}
\hypersetup{pdfauthor=Vladimír Holý and Michal Černý}

\usepackage{color}
\hypersetup{colorlinks=true, linkcolor=blue, anchorcolor=blue, citecolor=blue, filecolor=blue, urlcolor=blue}

\usepackage[margin=60pt]{geometry}
\usepackage{amsmath}
\usepackage{amssymb}
\usepackage{bm}

\usepackage[group-minimum-digits=3]{siunitx}

\usepackage{enumitem}

\usepackage{graphicx}
\usepackage{float}
\usepackage{hhline}
\usepackage{multirow}
\usepackage{rotating}
\usepackage{booktabs}

\usepackage[authoryear]{natbib}

\newtheorem{assumption}{Assumption}
\newtheorem{theorem}{Theorem}

\newcommand{\aA}{\mathfrak{A}}
\newcommand{\aB}{\mathfrak{B}}
\newcommand{\EE}{\mathsf{E}}
\newcommand{\var}{\textsf{var}\,}

\begin{document}

\begin{center}
{\Large \bfseries Bertram's Pairs Trading Strategy with Bounded Risk}
\end{center}

\begin{center}
{\bfseries Vladimír Holý} \\
Prague University of Economics and Business \\
Winston Churchill Square 1938/4, 130 67 Prague 3, Czech Republic \\
\href{mailto:vladimir.holy@vse.cz}{vladimir.holy@vse.cz}
\end{center}

\begin{center}
{\bfseries Michal Černý} \\
Prague University of Economics and Business \\
Winston Churchill Square 1938/4, 130 67 Prague 3, Czech Republic \\
\href{mailto:cernym@vse.cz}{cernym@vse.cz}
\end{center}

\begin{center}
{\itshape \today}
\end{center}

\noindent
\textbf{Abstract:}
Finding Bertram's optimal trading strategy for a pair of cointegrated assets following the Ornstein--Uhlenbeck price difference process can be formulated as an unconstrained convex optimization problem for maximization of expected profit per unit of time. This model is generalized to the form where the riskiness of profit, measured by its per-time-unit volatility, is controlled (e.g. in case of existence of limits on riskiness of trading strategies imposed by regulatory bodies). The resulting optimization problem need not be convex. In spite of this undesirable fact, it is demonstrated that the problem is still efficiently solvable. In addition, the problem that parameters of the price difference process are never known exactly and are imprecisely estimated from an observed finite sample is investigated (recalling that this problem is critical for practice). It is shown how the imprecision affects the optimal trading strategy by quantification of the loss caused by the imprecise estimate compared to a theoretical trader knowing the parameters exactly. The main results focus on the geometric and optimization-theoretic viewpoint of the risk-bounded trading strategy and the imprecision resulting from the statistical estimates.
\\

\\


\section{Introduction}

Pairs trading is a market-neutral trading strategy which exploits a long-term balance between two assets and makes profit when they are temporarily out of balance. The strategy assumes that the spread between the two assets is a stationary process implying mean reversion in finite time. It simply suffices to wait until the spread process is ``far'' enough from its mean value and then to bet it will revert to the mean by opening a long position in the underperforming asset and a short position in the overperforming asset. When the spread process indeed returns close to its mean, the positions are closed and the profit is made. In the traditional sense, this is a kind of ``free-lunch'', or arbitrage trading, where the profit is guaranteed. The tricky point is that mean reversion can be slow, meaning that waiting times for the collection of the almost sure profit can be long. Intuitively, it could be hardly said that there is a ``free-lunch'' profit once it is necessary to wait for it for an extremely long time period. 

There is a significant body of financial literature dealing with the pairs trading strategy. The main methods include the distance approach of \cite{Gatev2006a}, the cointegration approach of \cite{Vidyamurthy2004a}, the stochastic spread approach of \cite{Elliott2005}, the stochastic control approach of \cite{Jurek2007}, the machine learning approach of \cite{Huck2009}, the copula approach of \cite{Liew2013}, the principal components analysis approach of \cite{Avellaneda2010}, and the Hurst exponent approach of \cite{Ramos-Requena2017}. For a literature review on pairs trading, see \cite{Krauss2017}.

In this text, the stochastic spread approach is followed with focus on determining the optimal values of thresholds controlling the course of the strategy. \cite{Elliott2005} proposed to model the spread between the two assets in a pair as the Ornstein--Uhlenbeck process which is a stationary Gauss--Markov process in continuous time. \cite{Bertram2009,Bertram2010} then suggested to find the optimal entry and exit thresholds of the spread by maximizing the expected profit per unit of time. This can be formulated as an unconstrained optimization problem. Bertram's approach is further utilized by \cite{Cummins2012}, \cite{Zeng2014}, \cite{Goncu2016}, \cite{Endres2019b}, and \cite{Holy2018d}. Other works on the stochastic spread approach include \cite{Larsson2013}, \cite{Liu2017}, \cite{Bai2018}, \cite{Stubinger2018}, and \cite{Endres2019a}.

The work of \cite{Bertram2009,Bertram2010} is generalized by taking into account the variance of profit, resulting in a version augmented by a constraint formalizing the assumption that there exists an exogenous limit on maximal admissible riskiness (e.g., imposed by a regulatory authority). This constraint can decrease the expected per-time-unit profit, but it also decreases the probability that the profit-collection time would be extremely long. The idea of pairs trading with bounded risk was originally hinted by \cite{Holy2018d} while the current paper offers an in-depth treatment of the model including the impact of misspecification of the spread process.

\section{Formalization of the Strategy}

\begin{assumption}[price processes]
\label{ass:price}
\begin{itemize}
\item[(a)] There are two assets $\aA$ and $\aB$, with zero risk-free yields, the prices of which are driven by continuous-time price processes $A_s, B_s$; $s \geq 0$.
\item[(b)] There exists a constant $\eta$, called \textbf{cointegration coefficient}, such that the \textbf{price difference process}
$$
X_s := A_s - \eta B_s
$$
is stationary. 
\end{itemize}
\end{assumption}

\begin{assumption}[trading environment]
\label{ass:trade}
\begin{itemize}
\item[(a)] An investor is allowed to open positions in the assets $\aA, \aB$ both long and short.
\item[(b)] Trading can be performed in continuous time at arbitrary volumes (in particular, the assets are divisible).
\item[(c)] The num\'eraire is cash, which can be held long or short freely with zero yield/cost.
\item[(d)] There is an amount $c_{\aA} > 0$ ($c_{\aB} > 0$, respectively) called \textbf{transaction cost} per unit of asset $\aA$ (asset $\aB$, respectively), which is paid for an adjustment of a position in asset $\aA$ (asset $\aB$, respectively) by one unit.
\end{itemize}
\end{assumption}

For example, a purchase of three units of $\aA$ is associated with transaction costs $3c_{\aA}$. Similarly, a sale of three units of $\aA$ also costs $3c_{\aA}$ dollars. In the sequel, it will be useful to introduce the shorthand
$$
\tilde{c} := 2c_{\aA} + 2\eta c_{\aB}
$$
and refer to $\tilde{c}$ as \emph{transaction cost} for short.

\textbf{Remark.} The assumptions \ref{ass:price} and \ref{ass:trade} can be reformulated to accomodate for transaction costs per dollar (or unit of any currency). In that case, the price difference process would have the form $X_s := \ln A_s - \eta \ln B_s$ and the sequel would be the same.

Depending on particular properties of $X_s$, there may exist many ``free-lunch'' trading strategies. For example, the process might admit this strategy: 
\begin{itemize}
\item[(i)] wait until time $s$ with $X_s + \tilde{c}/2 < \mu := \EE X_s$;
\item[(ii)] buy the portfolio 
$$
(1, -\eta) := (\text{1 unit of $\aA$ long},\ \text{$\eta$ units of $\aB$ short})
$$
and pay the cost $\tilde{c}/2$;
\item[(iii)] wait until time $s' > s$ such that $X_{s'} \geq \mu + \tilde{c}/2$;
\item[(iv)] buy the portfolio $(-1, \eta)$ to close the positions and pay the cost $\tilde{c}/2$.
\end{itemize}
The overall profit, collected in time $s'$, is $X_{s'} - X_s - \tilde{c} > 0$, and it is achieved in finite time.

To be able to derive more detailed results, we need a particular form of the price difference process.
\begin{assumption}[Ornstein--Uhlenbeck process] The price difference process is of the Ornstein--Uhlenbeck form following the equation
\begin{equation}
\label{eq:vasicek}
\mathrm{d}X_s = \tau(\mu - X_s) + \sigma \mathrm{d}W_s,
\end{equation}
where $\mu$ stands for the mean $\EE X_s$, $\sigma > 0$ stands for the volatility, $\tau > 0$ measures the speed of mean reversion and $W_s$ is the standard Wiener process. 
\end{assumption}

Recall that equation \eqref{eq:vasicek} has a solution
$$
X_s = X_0 + \mu(1-e^{-\tau s})
+ \sigma\int_{0}^s
e^{-\tau(s-t)}\,\text{d}W_t.
$$
In addition, when  $X_0 \sim N(\mu, \sigma^2/(2\tau))$ and $X_0$ is independent of $W_s$, the process is stationary. Recall that this process is utilized in finance frequently, e.g.\ in Vasicek's interest rate model.

\begin{assumption}[no need to handle errors from econometric estimates]
The constants $\eta$, $\mu$, $\tau$, $\sigma$ are known (and need not be estimated 
from observable finite-sample data $(s_i, A_{s_i}, B_{s_i})_{i = 1, \dots, N}$).
\end{assumption}

This is a usual assumption in portfolio management theory; recall, for instance, that Markowitz also assumes the knowledge of \emph{exact} mean returns and \emph{exact} covariances of assets to be included in a portfolio (and not their econometric estimates, such as sample means and sample covariances).

The knowledge of parameters allows us, without loss of generality, to perform the transformation
$$
Y_t = \sqrt{\frac{2\tau}{\sigma^2}}(X_s - \mu),\quad
t = \tau s,
$$
leading to the \emph{standardized price difference process} $Y_{t}$ with zero mean and unit volatility. The transaction cost in the reparametrized model is then
\begin{equation}
c = \sqrt{\frac{2\tau}{\sigma^2}} \tilde{c}.
\end{equation}

\textbf{Remark.} In Section~\ref{sect:misspecif} the assumption will be relaxed. The Section is devoted to a study how an estimate of parameters, suffering from statistical imprecision, affects the ``optimal'' pairs trading strategy and what is the cost of the error induced by the fact that from finite samples it is never possible to retrieve the values of parameters $\eta, \mu, \tau, \sigma$ with full precision.

\section{Maximization of Expected Profit}\label{sect:B}

Steps (ii) and (iv) in the above sample strategy trigger trades (opening/closing of positions). The question is what are the ``best'' levels of $Y_t$ to open, close or adjust positions. Let us formalize this question in terms of so-called 
\emph{Bertram's trading strategy}. Let $a > 0$ stand for the \emph{entry level} and $b < a$ for the \emph{exit level} in the following strategy. 
\begin{itemize}
\item[(i)] Start in time $0$ and wait until time $t_1 \geq 0$ when $Y_{t_1} = a$ (``entry''), buy portfolio $(-1,\eta)$ and pay cost $c/2$.
\item[(ii)] Wait until time $t'_1 > t_1$ when $Y_{t'_1} = b$ (``exit'') and buy portfolio $(2, -2\eta)$ (i.e., close the positions from (i) and open the opposite positions), collect trading profit $a - b$ and pay transaction cost $c$.
\item[(iii)] Wait until time $t_2 > t'_1$ when $Y_{t_2} = a$ (``entry'') and buy portfolio $(-2, 2\eta)$
(i.e., close the positions from (ii) and open the opposite positions), collect profit $a - b$, pay cost $c$ and iterate forever.
\end{itemize}

For the sake of simplicity it is assumed that $t_1 = 0$, meaning that there is no ``idle-time'' in (i). This simplification does not affect the behavior of profit in the limit $t \rightarrow \infty$, which will be studied in the sequel. For the same reason, the cost $c/2$ from (i) can be neglected as well. 

The time 
$$
T_i := t_{i+1} - t_i
$$
is referred to as \emph{trade cycle}. The profit 
\begin{equation}\label{eq:profit}
\pi := 2(a - b - c)
\end{equation}
per trade cycle is deterministic. What is random here is the time $T_i$ to collection of the profit at the end of a cycle. In Bertram's strategy, profit is measured \emph{per time unit}. If $N_t$ is the counting process for the 
number of trade cycles in the time window $[0, t]$, then 
$$
\Pi(a,b) \equiv \Pi := \lim_{t \rightarrow \infty} \frac{\pi \EE N_t}{t} 
$$
is the \emph{expected profit per time unit}. 

\begin{figure}[t]
\centering
\includegraphics{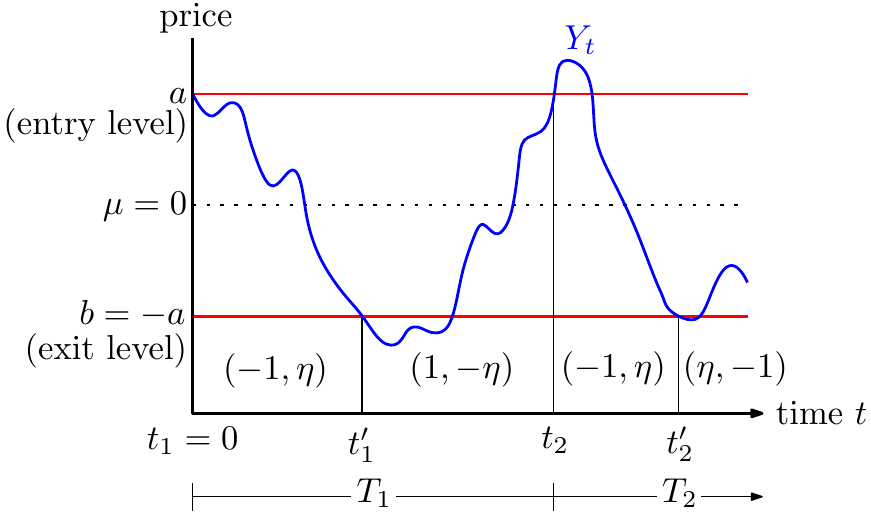}
\caption{Bertram's trading strategy: a sample trajectory of the price difference process $Y_t$, entry-exit levels $a,b$, trade cycles $T_1, T_2, \dots$ and portfolios $(-1, \eta)$ and $(1, -\eta)$ held in the corresponding cycles.}
\label{fig:strategy}
\end{figure}

\begin{theorem}
$$
\Pi(a,b) = \frac{\pi}{\EE T_i},
$$
where
$$
\EE T_i = \sum_{k=1}^{\infty}
\Gamma(k-\tfrac{1}{2})
\frac{(2^{1/2}a)^{2k-1} - (2^{1/2}b)^{2k-1}}{(2k-1)!}.
$$
\end{theorem}

Now the task is: given $c$ (still assuming standardized $Y_t$), 
solve
\begin{equation}\label{eq:original}
\max_{a,b} \Pi(a,b)\ \ \text{subject to}\ \ a \geq 0,\ a - b - c \geq 0.
\end{equation}
This is traditional \emph{Bertram's profit-maximization problem}. Observe that the constraint
$a - b - c \geq 0$ formalizes the natural assumption that strategies with profit at least zero are required.

The optimal solution $(a^*, b^*)$ is known to satisfy the property
$$b^* = -a^*,
$$
referred here to as \emph{symmetry} of the strategy (or symmetry of entry-exit thresholds).
The situation is depicted in Figure~\ref{fig:strategy}.

\begin{figure}[h]
\centering
\includegraphics{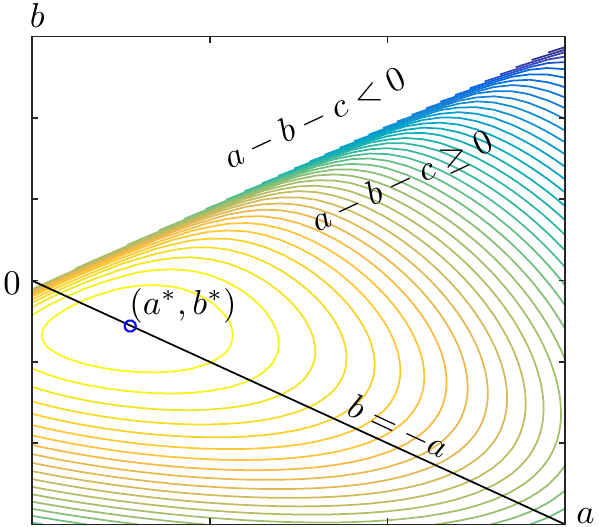}
\caption{Contour lines of expected profit-per-time-unit $\Pi$ as a function of the entry-exit thresholds $a, b$. The optimal thresholds, maximizing $\Pi$, are denoted by $(a^*, b^*)$. Here, it is assumed that the process $Y_t$ is standardized and that $c = 0.014$.}
\label{fig:ZMx}
\end{figure}

The optimal solution $(a^*, b^*)$ is not known explicitly (and possibly is not an elementary function of $c$), and thus it is obtained numerically. The geometry of the problem is depicted in Figure~\ref{fig:ZMx}.

\begin{figure}[h]
\centering
\includegraphics[width=\textwidth]{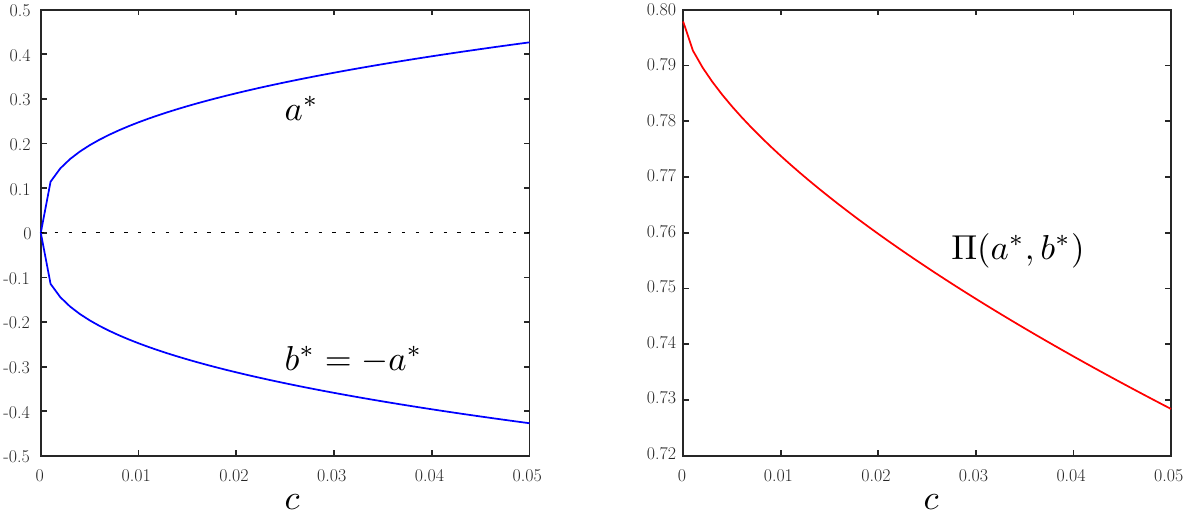}
\caption{The optimal trading strategy $(a^*, b^*)$ as a function of transaction costs $c$ and the resulting profit $\Pi(a^*, b^*)$.}
\label{fig:dep}
\end{figure}

\textbf{Remark.} It is interesting to study the dependence of the optimal trading strategy (entry-exit thresholds) $(a^*, b^* = -a^*)$ and the resulting expected profit $\Pi(a^*, b^*)$ as a function of transaction costs $c$. This is depicted in Figure~\ref{fig:dep}. Recall that wider gaps $a^* - b^*$ correspond to longer trade cycles $T_i$ and thus to less frequent trades (i.e., less frequently paid transaction costs $c$).

\section{Risk Bounds}
\label{sect:C}

\subsection{Variance of Profit per Time Unit as the Risk Measure}

Recall that the profit-per-time-unit $\Pi$ is subject to randomness through the random duration of a trade cycle. Thus, the variance of profit also reflects the variance of the length of a trade cycle $T_i$. Recall also that $N_t$ stands for the counting process for the number of trades in the time interval $[0, t]$. The overall profit over that period is 
$\pi N_t$ (where $\pi = 2(a - b - c)$ is the deterministic amount of profit per cycle). The variance of profit, standardized to a time unit, is then
\begin{equation}
V(a,b) \equiv V := \lim_{t \rightarrow \infty} \frac{\var(\pi N_t)}{t}.\label{eq:V}
\end{equation}
This is the risk measure to be controlled. (Separate section~\ref{sect:why} will be devoted to the question whether $t^{-1} \var(\pi N_t)$ is a `good' measure of risk in the long run $t \rightarrow \infty$.)

Assume that there is an exogenously given bound $v_0$ on the variance and that the task is to find a profit-maximizing strategy $(a,b)$ under the risk constraint $(\star)$:
\begin{equation}
\max_{a,b} \Pi(a,b) \text{\ \ subject to\ \ } 
a \geq 0,\  a - b - c \geq 0,\ 
\underbrace{V(a,b) \leq v_0}_{(\star)}.
\label{eq:constrained}
\end{equation}

\begin{figure}[t]
\centering
\includegraphics{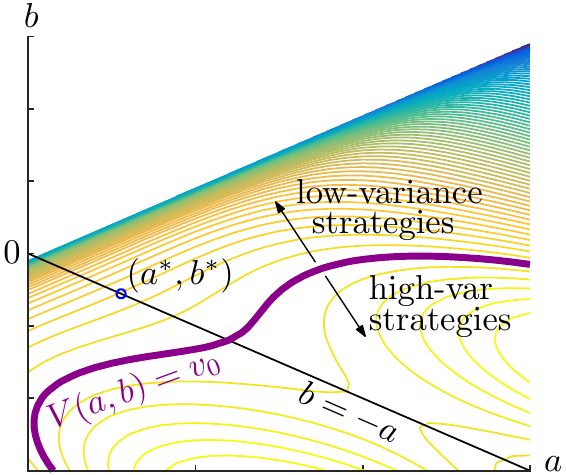}
\caption{Contour lines of variance (riskiness) $V(a,b)$ of a trading strategy (the pair of entry-exit thresholds) $a, b$. In this example the risk constraint $V(a,b) \leq v_0$ is inactive: the optimal solution $(a^*,b^*)$ of Bertram's unconstrained profit maximization problem is also an optimal solution of the risk-constrained problem.}
\label{fig:ZM}
\end{figure}

Let us illustrate the geometry behind (\ref{eq:constrained}). If $(a^*, b^*)$ is an optimal solution of  \eqref{eq:original} and $V(a^*, b^*) \leq v_0$, then the constraint $(\star)$ is redundant (inactive) and $(a^*, b^*)$ is also an optimal solution of (\ref{eq:constrained}). This situation is depicted in Figure~\ref{fig:ZM}. However, if $v_0$ is smaller, the constraint $(\star)$ might ``cut-off'' the point $(a^*, b^*)$ from the feasible region. Observe that the feasible region need not be convex in general, and thus (\ref{eq:constrained}) is not guaranteed to be a convex optimization problem. This issue will be elaborated on further in Section~\ref{sect:geometry}.

\subsection{Discussion on the Choice of the Risk Measure}\label{sect:why}

It is a natural question how to measure the risk of a strategy $(a,b)$. Profit per time unit is given by 
$\Pi_t := \pi N_t / t$. The objective function in (\ref{eq:constrained}) maximizes the expectation thereof in the long run $t \rightarrow \infty$. 
Thus it would be a natural choice to consider $\var \Pi_t$ as a risk measure and push it to the limit $t \rightarrow \infty$. However, 
$$
\var \Pi_t = \frac{\pi^2}{t^2}\var N_t \approx \frac{\pi^2}{t^2} \cdot \frac{t\cdot\var T_i}{(\EE T_i)^3} 
\stackrel{t \rightarrow \infty} \longrightarrow 0,
$$
since it is a known property of the Ornstein--Uhlenbeck process that $\var N_t \approx t (\EE T_i)^{-3} \var T_i$ for large $t$. Thus, $\lim_{t\rightarrow\infty}\var\Pi_t$ is a trivial risk measure. 

More generally, it is easy to see
that $\lim_{t\rightarrow\infty}\var(\pi N_t) / t^{\alpha}$ is a nontrivial function of $a,b$ only for $\alpha = 1$ 
(the limit is $0$ for $\alpha > 1$ and it is $\infty$ for $\alpha <1$).
This justifies why (\ref{eq:V}) is the `right' choice of the risk measure if it should reflect volatility and should be nontrivial in the long run (i.e., in the limit $t \rightarrow \infty$). 

\subsection{Geometry of the Risk-Bounded Trading Strategy}\label{sect:geometry}

Recall that $T_i$ is the duration of a trade cycle. From \cite{Bertram2010} it follows that:
\begin{theorem}
$$
V(a,b) = \frac{\pi^2}{(\EE T_i)^3}\var T_i,
$$
where
\begin{align*}
\var T_i &= w_1(a) - w_1(b) - w_2(a) + w_2(b), \\
w_1(\xi) &= 
\left( \frac{1}{2} \sum_{k=1}^\infty
\Gamma(\tfrac{k}{2})\frac{(2^{1/2}\xi)^k}{k!}
\right)^2
-
\left( \frac{1}{2} \sum_{k=1}^\infty
\Gamma(\tfrac{k}{2})\frac{(-2^{1/2}\xi)^k}{k!}
\right)^2, \\
w_2(\xi) &= \sum_{k=1}^\infty
\Gamma(k-\tfrac{1}{2})
\varphi(k-\tfrac{1}{2})
\frac{(2^{1/2}\xi)^{2k-1}}{(2k-1)!},
\end{align*}
and $\varphi(\cdot)$ is the digamma function.
\end{theorem}

\begin{figure}[htbp]
\centering
\includegraphics{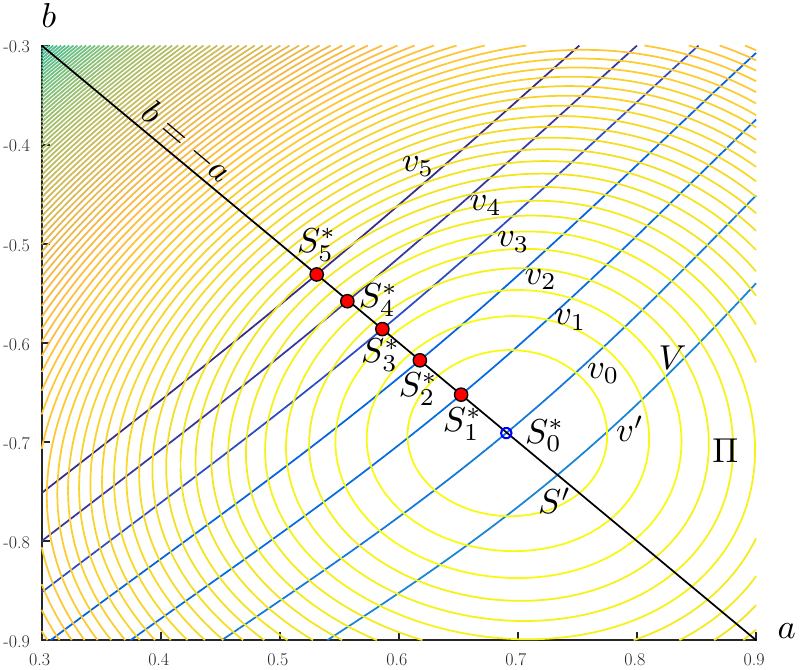}
\caption{Contour lines of the expected profit function $\Pi$ and variance $V$ on a neighborhood of the maximizer of $\Pi$ when $c = 0.2$. If the constraint $(\star)$ is in the form $V(a,b) \leq v'$, then the strategy $S^*_0 = (a^*, b^*)$ from (\ref{eq:original}) is optimal and the constraint $(\star)$ is redundant. The strategy $S'$ is never optimal. If the constraint $(\star)$ is $V(a,b) \leq v_i$ for an $i \in \{0, \dots, 5\}$, then the constraint $(\star)$ is active and the optimal strategy is $S_i^*$.} 
\label{fig:cns}
\end{figure}

The shape of contour lines of $V(a,b)$ in Figure~\ref{fig:ZM} are symmetric around the line $b = -a$. The profit function $\Pi(a,b)$ together with the contour lines of $V(a,b)$ --- the boundary of the feasible region constrained by $(\star)$ --- are plotted in Figure~\ref{fig:cns}. 

\begin{figure}[htbp]
\centering
\includegraphics{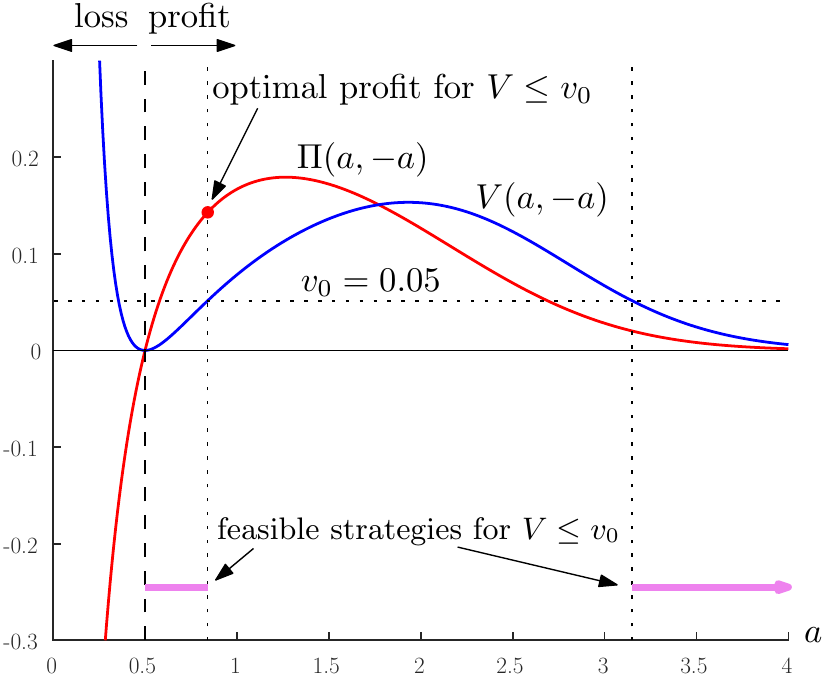}
\caption{Symmetric strategies with $c = 1$, the graph of profit $\Pi(a,-a)$ and variance $V(a,-a)$ as a function of $a > 0$. In $a = c/2 = 1/2$, both profit and its variance is zero. An example of the risk constraint $V \leq v_0 = 0.05$ and the optimal strategy.}
\label{fig:symm}
\end{figure}

The shape of contour lines from Figure~\ref{fig:cns} suggests that even in case (\ref{eq:constrained}), the symmetry condition is satisfied. Therefore it is possible to restrict the attention to the symmetric trading strategies $(a, b = -a)$ only. It is interesting to visualize such strategies in Figure~\ref{fig:symm}, where the profit $\Pi(a, -a)$ and variance $V(a, -a)$ are plotted as a function of~$a$. Observe that for $a = c/2$ it holds true that $\Pi = 0$, and thus $V = 0$. Moreover, the figure illustrates the interesting properties
\begin{align*}
\lim_{a \searrow 0} \Pi(a, -a) &= -\infty, & \lim_{a \rightarrow \infty} \Pi(a, -a) &= 0, \\
\lim_{a \searrow 0} V(a, -a) &= +\infty, & \lim_{a\rightarrow\infty} V(a, -a) &= 0.
\end{align*}

\begin{figure}[htbp]
\centering
\includegraphics{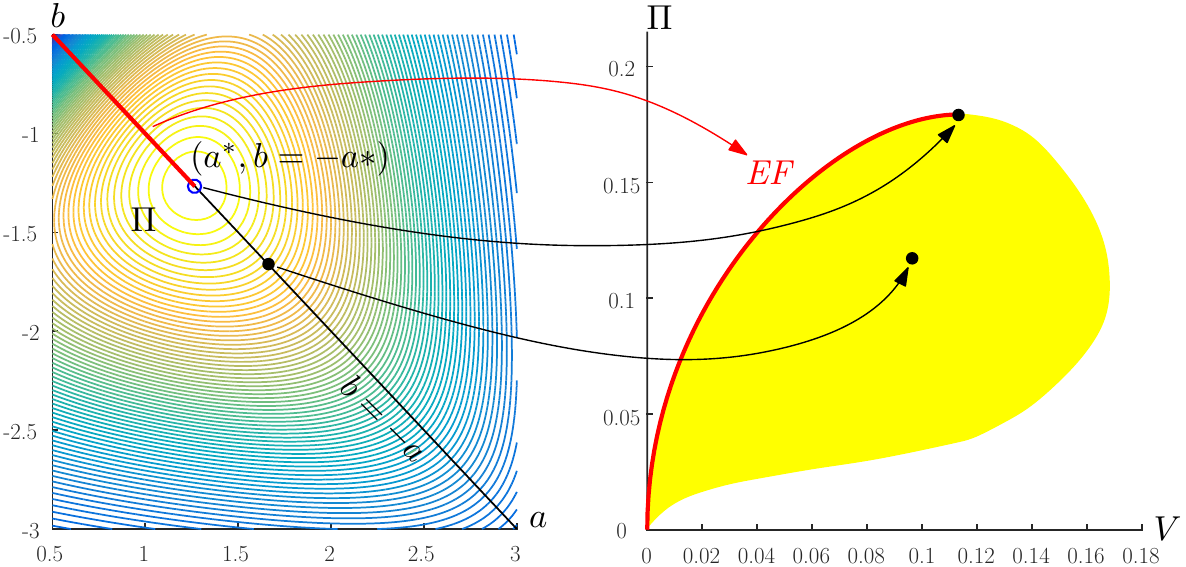}
\caption{Left subplot: the $(a,b)$-space of profitable strategies (satisfying $a - b - c \geq 0$ with $c = 1$)
and the profit function $\Pi$. Right subplot: the mean-var $(V,\Pi)$-space and an illustration how the space of strategies maps to the mean-var space. The efficient frontier (in red) is the image of $(a,-a)$ with $c/2 \leq a \leq a^*$.}
\label{fig:meanvar}
\end{figure}

Finally, Figure~\ref{fig:meanvar} shows how the space $(a,b)$ of strategies (not necessarily symmetric) maps to the mean-var space $(V, \Pi)$. More specifically, restricting only to the profitable strategies $a - b - c \geq 0$, the mean-var space visualizes the set of strategies $\{[V(a,b), \Pi(a,b)]\ |\ a - b - c \geq 0\}$ and the efficient frontier. Not surprisingly, the efficient frontier has the form 
\begin{equation}\label{eq:eff}
\textit{EF} := \left\{[V(a,-a), \Pi(a,-a)]\ \Big|\ \frac{c}{2} \leq a \leq a^*\right\},
\end{equation}
where $(a^*, -a^*)$ is the optimal solution of~(\ref{eq:original}).

\subsection{Solving the Risk-Constrained Problem}

By symmetry, the optimization problem (\ref{eq:constrained}) reduces to a problem in a single variable $a$, which can be solved easily by binary search (Algorithm 1). 
It is sufficient that the algorithm restricts the search space only to the strategies on the efficient frontier (\ref{eq:eff}), see also Figure~\ref{fig:meanvar}.

\vspace{.4cm}

\begin{tabular}{cl}
\hline
\multicolumn{2}{l}{\textbf{Algorithm 1.} Solving risk-constrained trading problem (\ref{eq:constrained})}\\
\hline
\{1\} & \textbf{Input:} cost $c > 0$, risk bound $v_0 > 0$, numerical tolerance $\varepsilon > 0$ \\
\{2\} & Solve the convex optimization problem (\ref{eq:original}) to find $(a^*, -a^*)$ \\
\{3\} & \textbf{If} $V(a^*, -a^*) \leq v_0$ \textbf{then} stop, $(a^*, -a^*)$ is optimal for (\ref{eq:constrained}) \\
\{4\} & $\overline{a} := a^*$, $\underline{a} := c/2$ \\
\{5\} & \textbf{While} $|V(\tfrac{1}{2}(\underline{a} + \overline{a}), -\tfrac{1}{2}(\underline{a} + \overline{a})) - v_0| > \varepsilon$
\textbf{do} \\
\{6\} & \ \ \ \ \ \textbf{If} $V(\tfrac{1}{2}(\underline{a} + \overline{a}), -\tfrac{1}{2}(\underline{a} + \overline{a})) > v_0$ \textbf{then}
        $\overline{a} := \tfrac{1}{2}(\underline{a} + \overline{a})$ \\
      & \hspace{5.43cm} \textbf{else} $\underline{a} := \tfrac{1}{2}(\underline{a} + \overline{a})$
        \\
\{7\} & \textbf{End while} \\
\{8\} & \textbf{Output} $\left(\tfrac{1}{2}(\underline{a} + \overline{a}), -\tfrac{1}{2}(\underline{a} + \overline{a})\right)$ as an $\varepsilon$-optimal strategy
for (\ref{eq:constrained})\\
\hline
\end{tabular}

\vspace{.4cm}

\section{Misspecification of the Price Difference Process}\label{sect:misspecif}

\subsection{Original Parametrization}

So far, the standardized price process $Y_t$ with zero mean and unit volatility has been dealt with. 
Now consider the general parametrization of the Ornstein--Uhlenbeck process  $X_s$ with parameters $\mu$, $\tau$ and $\sigma^2$. The entry and exit levels as well as transaction cost in general parametrization are obtained as
\begin{equation}
\tilde{a} = \sqrt{\frac{\sigma^2}{2\tau}} a + \mu, \qquad \tilde{b} = \sqrt{\frac{\sigma^2}{2\tau}} b + \mu, \qquad \tilde{c} = \sqrt{\frac{\sigma^2}{2\tau}} c.
\end{equation}
All three parameters therefore affect the entry and exit levels. Furthermore, the expected profit and the variance in general parametrization are obtained as
\begin{equation}
\tilde{\Pi}(\tilde{a}, \tilde{b}) = \sqrt{\frac{\tau \sigma^2}{2}} \Pi(a, b), \qquad \tilde{V}(\tilde{a}, \tilde{b}) = \frac{\sigma^2}{2} V(a, b).
\end{equation}

\subsection{Misspecification of $\mu$, $\tau$ and $\sigma^2$}

The effects of misspecified parameters of the price difference process are essential from the practical viewpoint. In theory, the knowledge of exact parameters can be assumed, but in practice misspecification occurs naturally whenever the parameters are statistically estimated from observed finite-sample data.

\begin{figure}
\centering
\includegraphics[width=\textwidth]{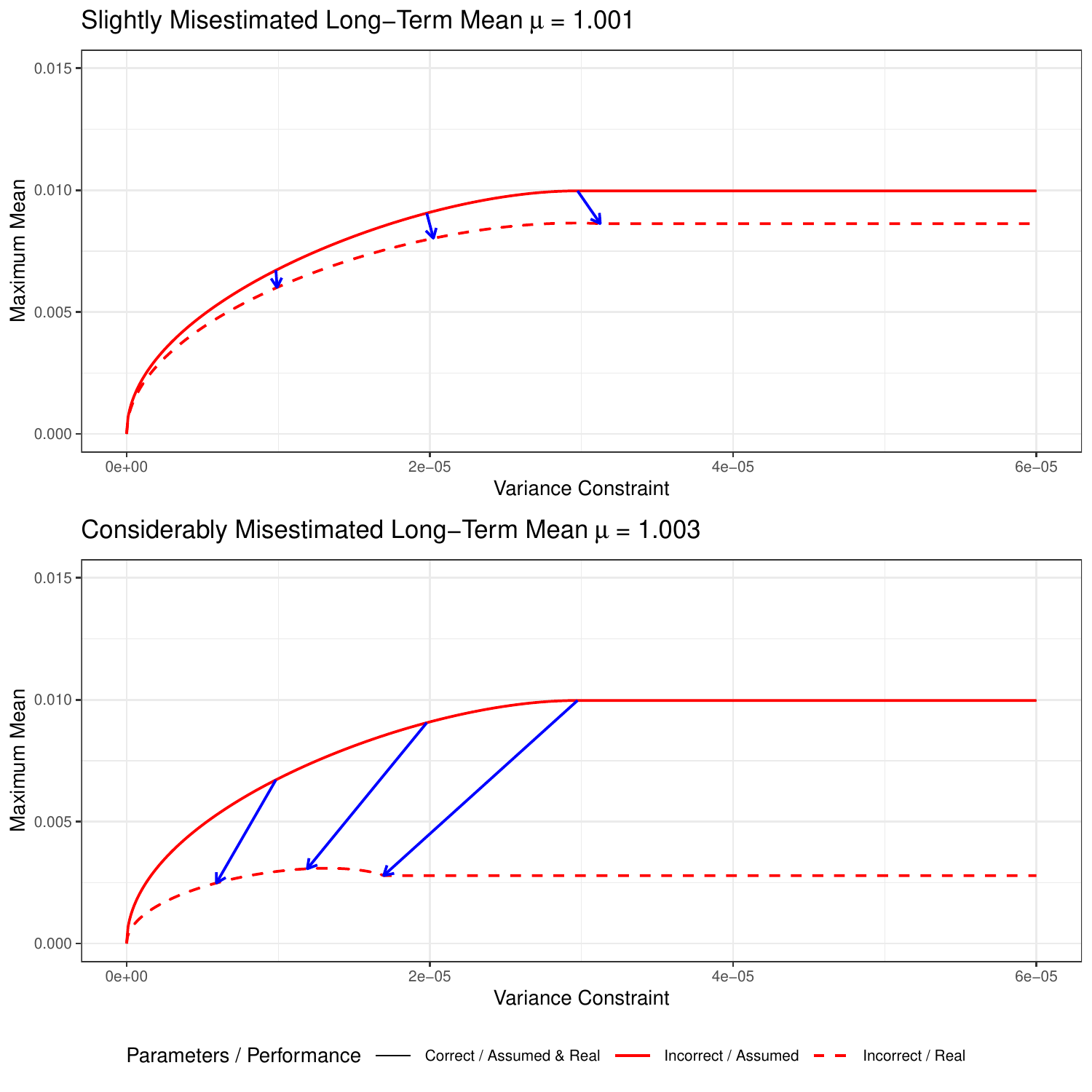}
\caption{The efficient frontier under misspecification of $\mu$ with true value $\mu = 1$.}
\label{fig:missMu}
\end{figure}

\begin{figure}
\centering
\includegraphics[width=\textwidth]{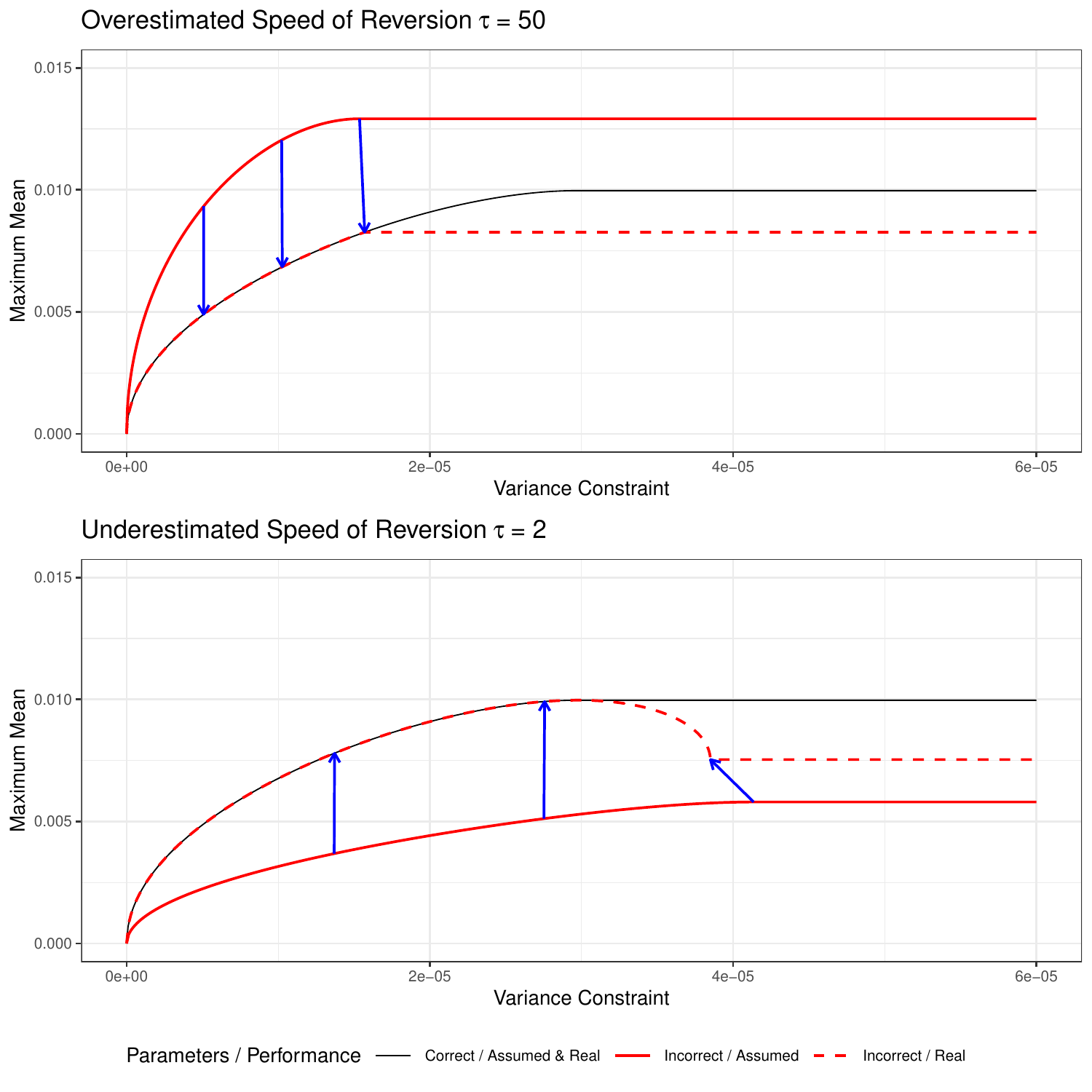}
\caption{The efficient frontier under misspecification of $\tau$ with true value $\tau = 10$.}
\label{fig:missTau}
\end{figure}

\begin{figure}
\centering
\includegraphics[width=\textwidth]{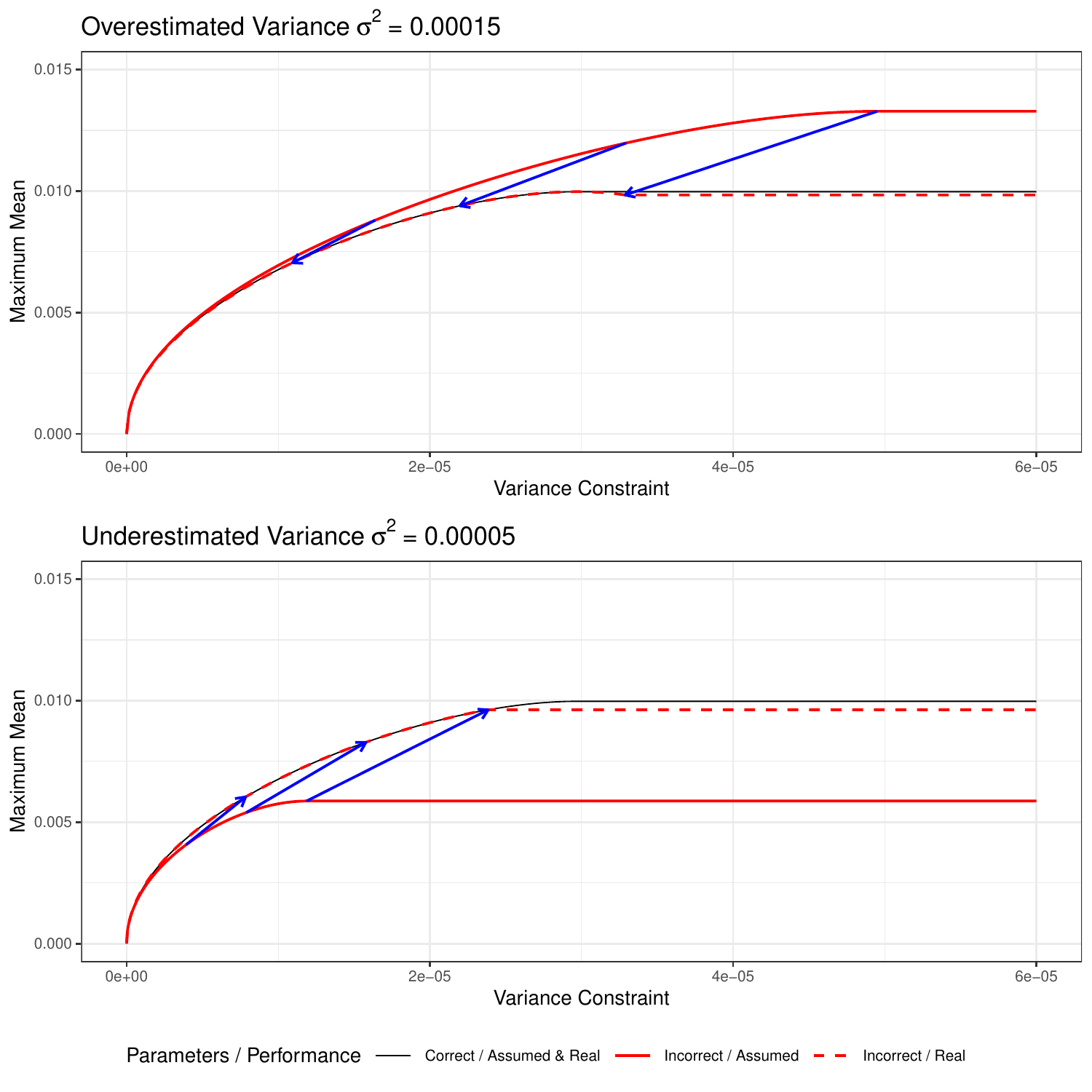}
\caption{The efficient frontier under misspecification of $\sigma^2$ with true value $\sigma^2 = 0.0001$.}
\label{fig:missSigma}
\end{figure}

Consider a benchmark case with $\mu=1$, $\tau=10$, and $\sigma^2 = 0.0001$ and set the transaction cost $\tilde{c}=0.0015$. This choice of the parameters is taken from \cite{Holy2018d} and reflects the values reported in their empirical study of the Big Oil companies. In figures \ref{fig:missMu}--\ref{fig:missSigma}, it is visualized how misspecification of these parameters affect the efficient frontier. When the long-term mean parameter $\mu$ is misspecified, the efficient frontier is believed to be exactly the same as for the correctly specified parameter. Misspecification, however, leads to suboptimal $\tilde{a}$ with lower expected profit and incorrect variance constraint.

When the speed of reversion $\tau$ is overestimated, the expected profit is believed to be higher than for the correctly specified parameter for any variance constraint. If the variance constraint in the misspecified model is binding, the expected profit is optimal but for larger variance constraint than desired. If the variance constraint is not binding, $\tilde{a}$ is lower than the unrestricted optimum and the expected profit is stuck at a suboptimal value. When the speed of reversion $\tau$ is understimated, the expected profit is believed to be lower than for the correctly specified parameter. If the variance constraint in the related correctly specified model is binding, the expected profit is optimal but for smaller variance constraint. If the variance constraint is not binding, $\tilde{a}$ is higher than the unrestricted optimum and the expected profit is suboptimal.

Finally, when the variance $\sigma^2$ is overestimated, the expected profit is believed to be higher than for the correctly specified parameter for any variance constraint. If the variance constraint in the related correctly specified model is binding, the expected profit is optimal but for smaller variance constraint than desired. If the variance constraint is not binding, $\tilde{a}$ is higher than the unrestricted optimum and the expected profit is suboptimal. When the variance $\sigma^2$ is underestimated, the expected profit is believed to be lower than for the correctly specified parameter. If the variance constraint in the misspecified model is binding, the expected profit is optimal but for larger variance constraint. If the variance constraint is not binding, $\tilde{a}$ is lower than the unrestricted optimum and the expected profit is stuck at a suboptimal value.

\subsection{Misspecification of $\eta$}

Recall that the existence of $\eta$, the cointegration coefficient, assures that $X_s$ is a stationary process. A wrong choice of $\eta$ might have devastating consequences, depending on the properties of the individual price processes $A_s, B_s$. It can happen that a wrong choice of $\eta$ can lead to non-stationarity of $X_s$. As a result, it could happen --- for example --- that the mean-reversion feature disappears and the trading strategy would lead to unbounded losses, either with nonzero probability, or even with probability tending to one. An example of a less serious (but still harmful) consequence is that that the trade cycles can be extremely long, say $\EE T_i \rightarrow \infty$ with $i \rightarrow \infty$. Then the speed of growth of $\EE T_i$ would be critical for practical considerations. And many more undesirable situations can occur. A detailed analysis of such phenomena deserves further investigation, based on an inspection of possible forms of the price processes $A_s, B_s$, namely in case when both of them are non-stationary.

\section{Conclusions}

As the first main result, the risk-constrained version of Bertram's trading strategy with a pair of cointegrated assets has been designed. Its geometry has been studied and there has also been proposed a solution method for finding the profit-maximizing strategy respecting the risk constraint. The second main result is the study of effects of misspecification of parameters of the underlying price difference process of the Ornstein--Uhlenbeck type. This is essential for practice since the parameters are never known exactly and are always estimated from a finite sample of observations.

In this model, risk has been measured by variance of profit normalized to a time unit. Recall that the variance of profit is driven by the variance of the length of a trade cycle. Further research should focus on other risk measures and the geometry of the optimization problem with multiple risk constraints (e.g., when both the average length of a trade cycle and its volatility are bounded). In addition, the trading strategy can be expected to be generalized to a wider class of stationary price processes. Finally, the profit $\pi$ per cycle has been considered in the form \eqref{eq:profit}. However, it would be suitable to treat $\pi(a,b,c)$ as a more general function modeling other constructions of transaction costs. For example, $\pi$ could take into account increasing costs of short selling over time. Short selling constraints could also be added to the model in the form of a maximum spread between the prices at which the trader would be forced to close both short and long positions at a loss.

\section*{Funding}
\label{sec:fund}

The work was supported by Czech Science Foundation under grant 19-02773S and the Internal Grant Agency of the Prague University of Economics and Business under grant F4/27/2020.


\end{document}